\documentclass[%
preprint, 
 amsmath,amssymb,
 aps,
]{revtex4-1}
\usepackage{graphicx}
\usepackage{dcolumn}
\usepackage{bm}
\usepackage{float}
\begin{document}
\title{Information processing via human soft tissue}
\author{Yo Kobayashi}
\email{kobayashi.yo.es@osaka-u.ac.jp}%
\affiliation{Graduate School of Engineering Science, Osaka University, Osaka, Japan}%
\date{\today}
\begin{abstract}
This  study demonstrates that the soft biological tissues of humans can be used as a type of soft body in physical reservoir computing. Soft biological tissues possess characteristics such as stress-strain nonlinearity and viscoelasticity that satisfy the requirements for physical reservoir computing, including nonlinearity and memory. The aim of this study was to utilize the dynamics of human soft tissues as a physical reservoir for the emulation of nonlinear dynamical systems. To demonstrate this concept, joint angle data during motion in the flexion-extension direction of the wrist joint, and ultrasound images of the muscles associated with that motion, were acquired from human participants. The input to the system was the angle of the wrist joint, while the deformation field within the muscle (obtained from ultrasound images) represented the state of the reservoir. The results indicate that the dynamics of soft tissue have a positive impact on the computational task of emulating nonlinear dynamical systems. This research suggests that the soft tissue of humans can be used as a potential computational resource.
\end{abstract}
\maketitle
\section{\label{sec:introduction}Introduction}
Reservoir computing is an information processing technique that exploits the input-driven transient behavior of a high-dimensional system known as a reservoir, and serves as a framework for recurrent neural network learning \cite{jaeger2002tutorial, maass2002real, goto2021twin}. Reservoir computing is particularly well-suited for real-time computing with time-varying inputs, and is therefore highly effective for emulating complex temporal computations \cite{maass2002real, jaeger2002tutorial, jaeger2004harnessing, buonomano2009state, verstraeten2007experimental,nakajima2015information}. A reservoir computing approach has been proposed to overcome the limitations of conventional neural networks, which take a considerable amount of time to converge to an attractor state and are not ideal for real-time computation \cite{nakajima2015information}. Complex nonlinear dynamical systems can be effectively emulated by incorporating linear and static readouts from the reservoir's high-dimensional state space, provided that the reservoir dynamics exhibit sufficient nonlinearity and memory.  This technique offers the advantage of maintaining low computational costs for both learning and computation, which is achieved by adjusting the linear and static readout weights from the reservoir. Various abstract dynamical systems have been proposed for reservoir applications, including neural models for echo state networks \cite{jaeger2002tutorial, jaeger2004harnessing} and liquid state machines \cite{maass2002real}, water flow states \cite{fernando2003pattern, goto2021twin}, and nonlinear mass spring systems \cite{hauser2011towards, hauser2012role}. 

A recent development in reservoir computing is the use of physical system dynamics as a reservoir, a technique referred to as physical reservoir computing \cite{tanaka2019recent, nakajima2020physical}. It has been suggested that a physical system with nonlinearity and memory could serve as an alternative to traditional computational frameworks like the Turing machine \cite{nakajima2014exploiting}. Examples of implementation have been reported in a variety of physics fields, including photonics \cite{vandoorne2008toward, larger2012photonic, brunner2013parallel}, spin \cite{torrejon2017neuromorphic, tsunegi2019physical}, and quantum \cite{fujii2017harnessing, nakajima2019boosting}, with each platform demonstrating particular computational properties intrinsic to its respective spatio-temporal scale \cite{goto2021twin}. These examples demonstrate that the suitability of a system for use as a reservoir is not contingent upon its specific implementation, but rather relies on system properties such as input separability and fading memory \cite{nakajima2015information}.

Several studies have demonstrated that soft body dynamics  can be used as a physical reservoir \cite{nakajima2015information, nakajima2014exploiting, nakajima2018exploiting}. In these studies, the mechanical system of a soft body composed of a rubber material, designed to mimic the legs of an octopus operating underwater, functions as a reservoir. The dynamics resulting from the interaction between the soft body and the water acted as a reservoir, confirming that the viscous damping caused by the water was necessary to preserve the memory for the calculations \cite{kagaya2022echo}. The kinematics and material nonlinearity of the soft body were also utilized to act as a reservoir \cite{nakajima2013soft}. 

This study demonstrates that the soft biological tissue of humans can be utilized as a physical reservoir for information processing. Soft biological tissues exhibit stress-strain nonlinearity and possess viscoelastic properties \cite{fung2013biomechanics, kobayashi2017simple, kobayashi2020non}. Therefore, the dynamics of soft tissue satisfy the system requisites, including nonlinearity and memory, necessary for physical reservoir computing. The specific objective of this study is to utilize the dynamics of soft human tissues as a physical reservoir to emulate nonlinear dynamical systems. Soft tissues are found throughout the human body; therefore, if computational processing can be delegated to biological tissues, it could lead to a distributed computation system for human-assisted devices. To the best of our knowledge, this is the first study to discuss the capability of \textit{in vivo} human tissue to act as a physical reservoir.

\begin{figure*}
\includegraphics[width=12cm]{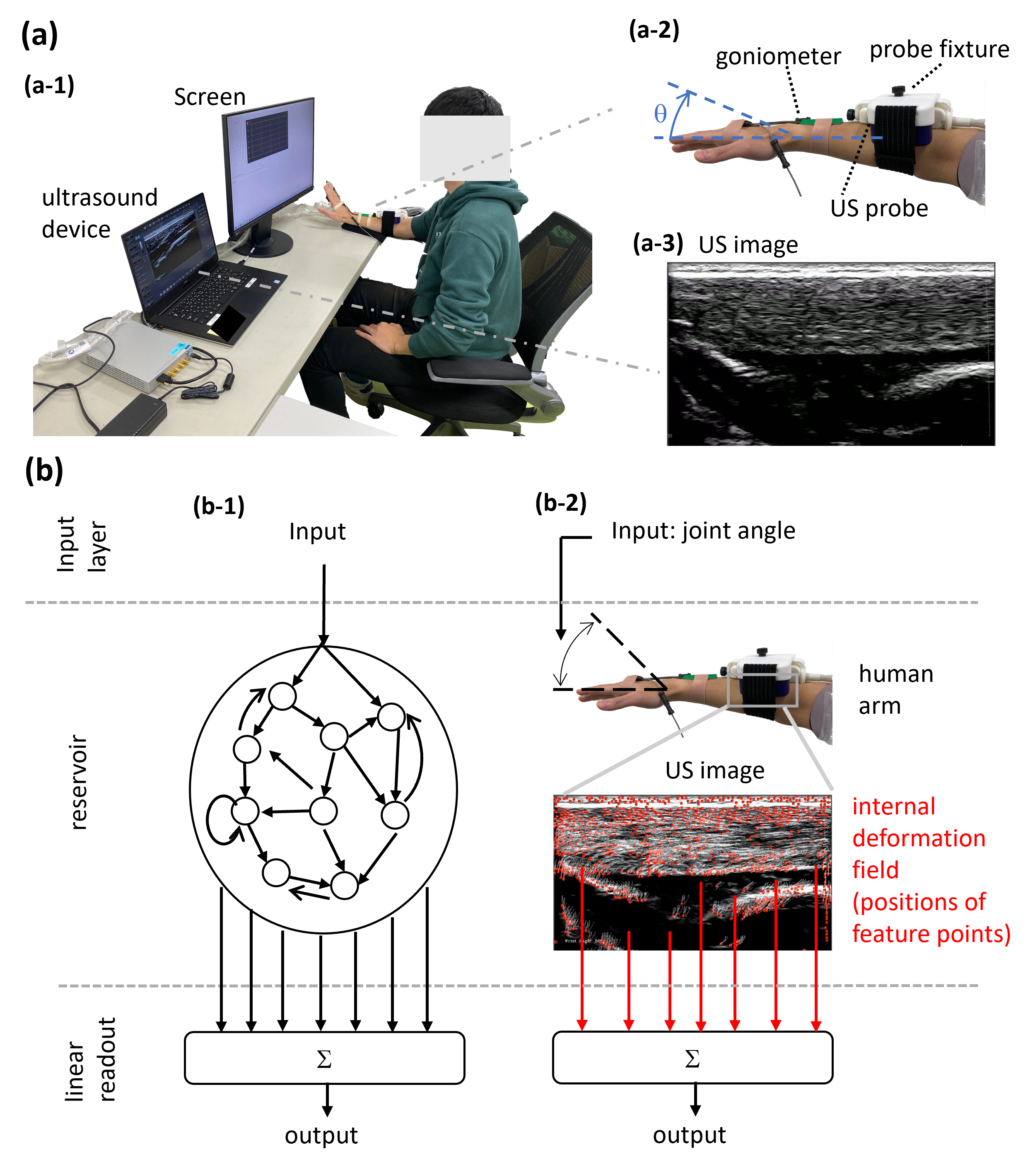}
\caption{\label{fig:concept}
Experimental setup and information processing scheme using human soft tissue. \textbf{(a)} Photograph illustrating the experimental setup employed in this study. The wrist joint angle (a-2) and ultrasound images (a-3) of the muscles associated with the movement in the flexion-extension direction of the wrist joint were acquired from human participants (a-1).  The objective of this experiment was to generate muscle deformation dynamics though wrist joint movement and to use the deformation as a physical reservoir for information processing. \textbf{(b)} Schematic representation of an analogy between a conventional reservoir computing system (b-1) and our system (b-2) used in this study. In conventional reservoir systems, randomly coupled abstract computational units are used as reservoirs (b-1). In our system, the deformation fields (positions of feature points) inside the muscle, obtained from image processing of ultrasound images, are used as computational units (b-2).  The focus of this study is to investigate if this physical reservoir can effectively emulate nonlinear dynamical systems often targeted by conventional reservoir computing systems.
}
\end{figure*}
\section{\label{methods} Methods}
The experimental concept for this study is illustrated in Fig.\ \ref{fig:concept}. The joint angle during motion in the flexion-extension direction of the wrist joint and ultrasound images (echo movies) of the muscles associated with that motion were acquired from human participants. This experiment aims to generate muscle deformation dynamics by varying the wrist joint angle and utilize the resulting deformation field as a physical reservoir for information processing. The input to the system is the wrist joint angle, and the state variable of the reservoir is the deformation field within the muscle. To successfully emulate nonlinear dynamical systems within this setup, the dynamics of the soft tissue must provide the necessary nonlinearity and memory, because reservoir computing relies solely on static and linear readouts during the learning process.  

Data were collected from 10 young, healthy participants aged 20 years or older, who had completed their growth spurts. This research was approved by the ethical committee for human studies at the Graduate School of Engineering Science, Osaka University. The experimental procedure was identical for all participants. The experimental setup is shown in Fig.\ \ref{fig:concept} (a). A goniometer (SG110/A, Biometrics Ltd) was attached to the wrist joint. The probe (LF11-5H60-A3, Telemed Medical Systems) of the ultrasound imaging device (ArtUS EXT-1H, Telemed Medical Systems) was attached to the position where a large muscle deformation was observed near the elbow joint during flexion and extension of the wrist joint, and was secured in place with a band (ProbeFix Dynamic T, Telemed Medical Systems). The ultrasound probe was positioned to extensor carpi radialis brevis muscle and its surrounding muscles (approximately 5[cm] from the elbow joint toward the wrist joint). The extensor carpi radialis brevis muscle is a muscle that contributes to flexion and extension of the wrist joint. 

Participants performed wrist joint extension motion along the target line, displayed on screen as the target joint angle. A sinusoidal curve was selected to specify the target joint angle (ref: Fig.\ \ref{fig:time}(a)), with 0 [deg] used as a reference position, and either 20[deg], 40[deg], or 60[deg] randomly selected as the amplitude of the sinusoidal curve. The period of the target joint angle was 4 seconds. During each experiment, 18 cycles of the specified motion was measured. Taking into account that noise was likely to occur at the beginning and end of each experiment, the data from 15 cycles were used for analysis (the initial two cycles and final cycle of each data set was removed). The data used in the evaluation phase described below included joint angles of all amplitudes (20[°], 40[°], and 60[°]) (ref: Fig.\ \ref{fig:time}(a)). Five sets of experiments were performed per participant, resulting in a total of 50 experiments from the 10 participants. 

Time series data of the wrist joint angles acquired during the experiment were used to calculate the input joint angle vector \textit{$ \boldsymbol\theta= \left (\theta_{1}, \theta_{2}, \cdots   \right )$} where \textit{$\theta_{k}$} is the joint angle at the k-th time data point. Ultrasound images of the muscle were acquired and feature points were extracted by image processing at each time point. To determine the initial positions of the feature points, Shi-Tomasi corner detection \cite{shi1994good} was used for the first frame of the ultrasound image. The coordinates of each feature point extracted by corner detection were tracked using optical flow \cite{lucas1981iterative} to obtain their x, y coordinate data \textit{$ \mathbf{\bar{s}} = \left (\bar{x}^{1},\bar{y}^{1}, \bar{x}^{2}, \bar{y}^{2}, \cdots   \right )^T $}, where (\textit{$\bar{x}^{i}$}, \textit{$\bar{y}^{i}$} ) is the coordinate of the i-th feature point. These data were used to calculate the reservoir state vector \textit{$ \mathbf{s_k} = \left (s^{1}_k,s^{2}_k, s^{3}_k, s^{4}_k, \cdots   \right )^T$} at each time point  k. See Appendix \ref{Preprocess} for details of this data preprocessing.

To evaluate the computational capability of our system, we used a nonlinear autoregressive moving average (NARMA) model as an emulation task for nonlinear dynamical systems, which are the standard benchmark tasks in the field of machine learning, in particular for recurrent neural network learning. This task presents a challenging problem for any computational system due to its dependence on nonlinearity and long time lags \cite{hochreiter1997long}. The first NARMA system is the following second-order nonlinear dynamical system. This system was introduced in \cite{atiya2000new} and was used as an example in \cite{hauser2011towards, nakajima2013soft, nakajima2015information, goto2021twin}. Hereafter we refer to this system as NARMA2.
\begin{eqnarray}
o_{k+1}=0.4 o_k+0.4 o_k o_{k-1}+0.6(r_{k})^3+0.1
\label{eq:1}
\end{eqnarray}
where \textit{$r_{k}$} and \textit{$o_{k}$} are input and system output at k-th time point. The next NARMA system is a nonlinear dynamical system of order N, represented as follows:
\begin{eqnarray}
o_{k+1}=0.3 o_k+0.05 o_k\left(\sum_{j=1}^{N-1} o_{k-j}\right)+1.5 r_{k-N+1} r_k+0.1
\label{eq:2}
\end{eqnarray}
The value of N ranges from 3 to 15, and each corresponding system is denoted as NARMA N (e.g., when N = 10, it is referred to as NARMA10). Notably, NARMA10 with this particular parameter setting was first introduced in \cite{atiya2000new} and has been widely used in various studies, such as \cite{verstraeten2007experimental, hauser2011towards, nakajima2013soft, nakajima2015information, goto2021twin}. The aim of these emulation tasks is to test the system's ability to process input histories with nonlinearity. A key aspect of the NARMA model is to include the long-term dependence of the system with an N-th time lag, which requires that the learning system have some degree of nonlinearity and memory to emulate it \cite{goto2021twin}. The input vector \textit{$ \mathbf{r} =  (r_{1},r_{2}, \cdots ) $} was calculated from the input joint angle vector \textit{$\boldsymbol \theta$} using the equation \textit{$r_k= 0.2 \times \theta _{k}/60$} to scale linearly to the approximate range [0, 0.2], based on previous studies \cite{goto2021twin}.

Thus, we applied this input \textit{$r_k$} to NARMA2 through NARAMA15 models and obtained the target output \textit{$o_k$}. The readout was implemented using a linear model commonly used in reservoir computing. The system output \textit{$ \mathbf{\hat{o}} =  (\hat{o}_{1}, \hat{o}_{2}, \cdots) $} is then defined as follows:
\begin{eqnarray}
\hat{o}_k = \sum_{i} w^i  s^i_k
\label{eq:4}
\end{eqnarray}

To emulate the desired nonlinear dynamical system in this framework, we adjust only the linear readout weights \textit{$\mathbf{w}=(w^{1},w^{2}, \cdots)$}, which are fixed after learning. Ridge regression \cite{hastie2009elements} was used to learn these weights \textit{$\mathbf{w}$}. See Appendix \ref{RidgeRegression} for details on the learning process using ridge regression. 

A total of 15 cycles (60 seconds) of data were used for the analysis, which was divided into a training phase and an evaluation phase. The training phase consisted of the first 12 cycles (48 seconds) and the evaluation phase encompassed the remaining 3 cycles (12 seconds). The first 1.5 seconds of the training data was omitted from the training dataset because it  contained the initial transient response of the nonlinear equations. During the training phase, linear readouts were trained for each task using the aforementioned ridge regression procedure. In the evaluation phase, trained linear readouts were used to generate system outputs \textit{$ \mathbf{\hat{o}}$}. 

During the evaluation phase of this study, the normalized mean square error (NMSE \textit{$=\sum_{k}\left(o_k-\hat{o}_k\right)^2 / \sum_{k}\left(o_k\right)^2 $} ) was calculated for the target and system outputs to assess the performance of the system output. For each of the 50 experimental data runs, the NMSE was calculated for each of the NARMA2 to NARMA15 emulation tasks. From these results, we calculated the means and variances of the NMSE. 

To characterize biological soft tissue as a physical reservoir, we also computed task performance using a simple linear regression (LR) model \cite{nakajima2015information, goto2021twin}. A simple LR model is represented by the equation \textit{$\hat{o}^{LR}_k = w'' \times r_k + w'$}. It is important to note that this model assumes the absence of soft tissue dynamics and only considers the raw inputs for information processing; therefore,  we can conclude that soft tissue contributes to the emulation task of the nonlinear equation if our system demonstrates superior computational performance when compared with the simple LR model. In the accuracy validation of these simple LR models , the training data, training method, and training procedure were the same for those used in our system, to ensure fair comparison.

\section{\label{result} Result}
\begin{figure*}
\includegraphics[width=12cm]{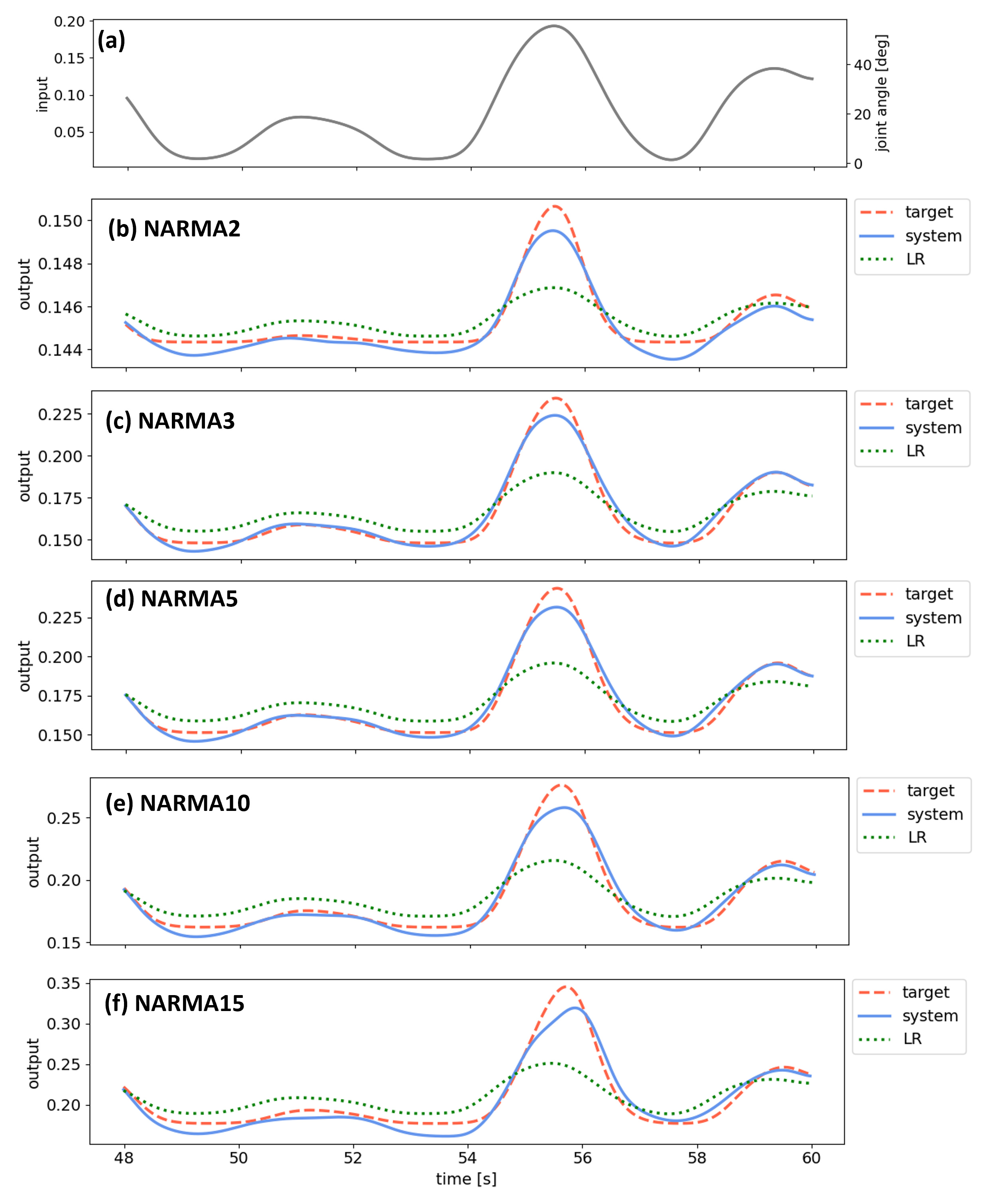}
\caption{\label{fig:time} 
Representative examples of time series data of the input and outputs from the evaluation phase. Time series data are presented for (a) the input (and joint angle) and the outputs for (b) NARMA2, (c) NARMA3, (d) NARMA5, (e) NARMA10, and (f) NARMA15. The plots for each output (b)-(f) are overlaid with the time series of the target output \textit{$ \mathbf{o}$}  (dashed red line), the system output \textit{$ \mathbf{\hat{o}}$} (blue line), and the output of the simple LR model \textit{$\hat{o}^{LR}$} (doted green line) for comparison. The system outputs demonstrate superior performance compared with the simple LR model across all NARMA tasks. These results indicate that the dynamics of human soft tissue play a significant role in emulating the nonlinear equations.
} 
\end{figure*}

Representative examples of the performance of the inputs and the NARMA2, 3, 5, 10, and 15 tasks during the evaluation phase are presented in Fig.\ \ref{fig:time}, along with the performance of the simple LR model for comparison purposes. It is evident that our system outperforms the simple LR model in all NARMA tasks. The emulation results using our system closely approximate the target dynamics. In contrast, the simple LR models based on raw inputs, which do not account for the dynamics of soft tissue, fail to emulate the target dynamical system.  These results suggest that the dynamics of soft tissue contribute to the successful emulation of the nonlinear equation. 

\begin{figure*}
\includegraphics[width=12cm]{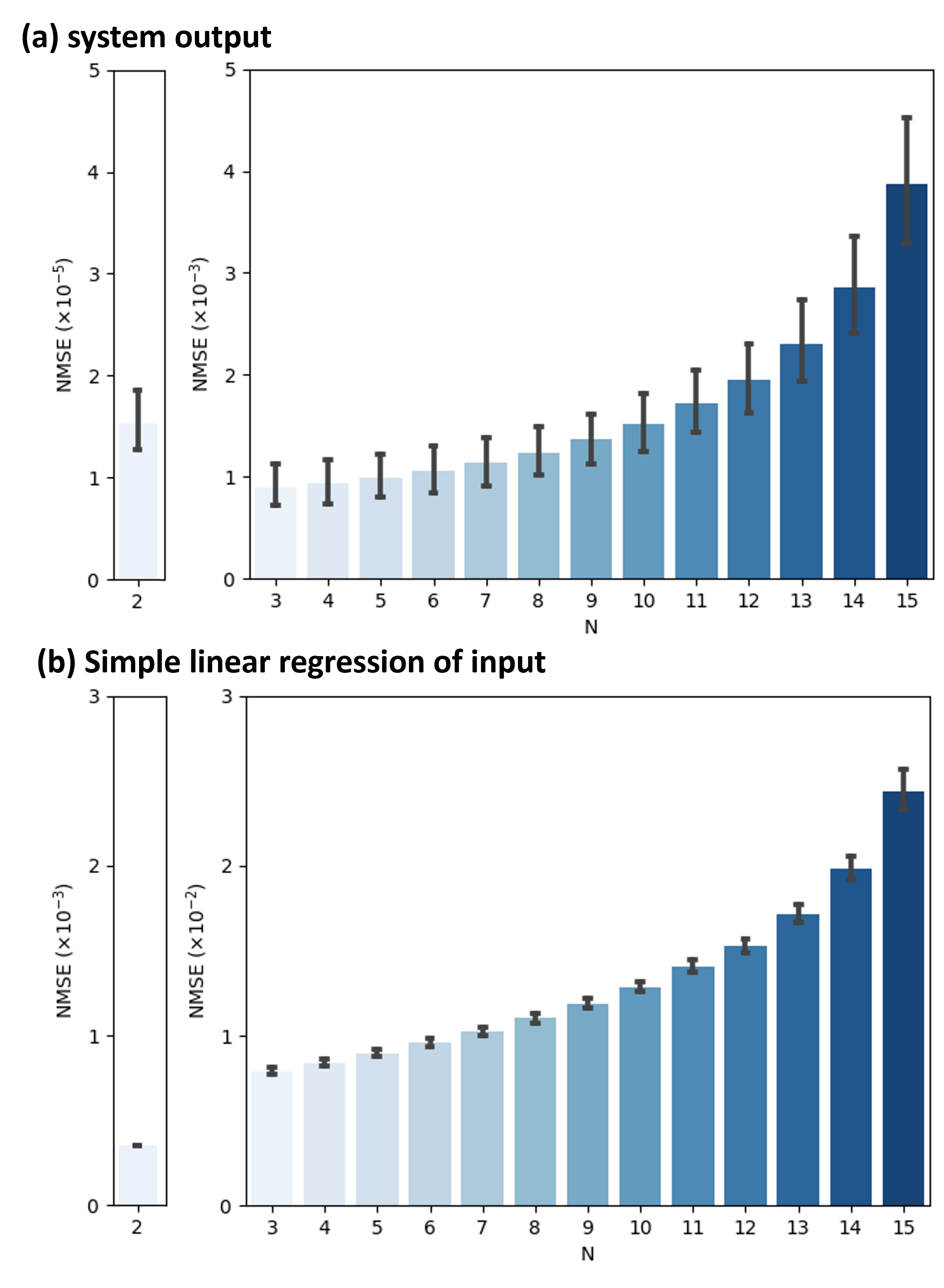}
\caption{\label{fig:stats} 
Comparison of the mean error and mean variation of the NMSE for each NARMA task. \textbf{(a)} The NMSE of the system output \textit{$ \mathbf{\hat{o}}$} using the dynamics of human soft tissue. \textbf{(b)} The NMSE of the simple linear regression model of the input \textit{$\hat{o}^{LR}$}. The horizontal axis represents the order N of the NARMA model.  Note that the vertical axes of (a) and (b) have different scales. For all the NARMA tasks, the NMSE of our system is lower than that of the simple LR model. These results demonstrate the potential for using human soft tissue as a computational resource.
} 
\end{figure*}

The mean and variance of the NMSE for the NARMA tasks at each N are presented in Figure\ \ref{fig:stats}. Figure\ \ref{fig:stats} (a) presents the results of our system incorporating the dynamics of soft tissues, while Fig.\ \ref{fig:stats} (b) presents the results of the simple LR model. In all NARMA tasks, the NMSE values achieved by our system were consistently lower than those of the simple LR model. A statistical comparison using the paired t-test between our system and the simple LR model reveals significant differences (\textit{$ p<0.01$}) for all NARMA tests. The performance of our system gradually decreases as N increases (Fig.\ \ref{fig:stats}(a)), indicating a limitation in the memory capacity of the soft tissue dynamics. 

These results demonstrate that our system successfully performs its task by exploiting the nonlinearity and memory derived from the dynamics of soft tissues.

\section{\label{discussion} Discussion}
In this study, we demonstrated that the dynamics of soft biological tissue can be exploited to emulate nonlinear dynamical systems. A comparative analysis of task performance with the simple LR model established the superior performance of our system across all tasks. These results demonstrate the positive impact of soft tissue dynamics on the computational task.  Moreover, these results suggest that human soft tissue can potentially be used as a computational resource. In other words, with appropriate input sets and computational tasks defined, soft biological tissue can be utilized as part of a computational device. Unlike other conventional computers based on the Turing machine, our system does not require an explicit storage device and assumes that memory is implicit in the transient dynamics of soft tissues. By assigning tasks to the system that require memory (NARMA tasks) and evaluating its performance, we could confirm the memory capacity of our system. In this study, joint motion was used as the method to induce muscle deformation, but external actuator devices could also be used in such applications. Since biological soft tissue is present throughout the human body, this system has the potential to eliminate or reduce the computational requirements for devices in close proximity to humans by delegating the computational load to human soft tissue. 

Our approach is capable of simultaneously emulating multiple nonlinear dynamical systems (NARMA2 to NARMA15) using the dynamics of the same soft tissue by adjusting the weight of the read out. This means that the physical reservoir of soft tissues (the single data series of the deformation field \textit{$\mathbf{S}$}) can compute multiple tasks simultaneously, a concept known as multitasking \cite{hauser2011towards, nakajima2014exploiting}. This suggests that when soft tissue is used as a physical reservoir, it is not necessary to have a different physical system for each computational task; instead, a single physical system can perform multiple computational tasks. 

In reservoir computing research, it has been established that for a reservoir to have strong computational power, it needs to possess input separability and fading memory \cite{maass2002real, nakajima2014exploiting}. In the case of a soft tissue physical reservoir, it is believed that these properties are inherent in its dynamics because of the viscoelasticity and nonlinear stress-strain properties of soft biological tissue. Input separability is usually accomplished by nonlinear mapping from low-dimensional inputs to higher dimensional state spaces. The body dynamics of soft materials tend to be underactuated systems \cite{tedrake2009underactuated}, which would allow input signals to naturally map to higher dimensions of the soft body, thus achieving input separability \cite{nakajima2014exploiting}. The non-uniformity and diverse mechanical properties of soft biological tissues potentially facilitate input separability. Fading memory is a property that preserves the effects of recent input sequences in the system, allowing the integration of input information over time. This ensures a reproducible computation, where the recent history of the input signal is important. In our system, the slow relaxation of the dynamics to the input signal, due to the damping effect caused by the viscoelastic properties of soft tissue, is believed to help achieve fading memory.  Previous studies on physical reservoirs using rubber materials for soft bodies operating in water \cite{nakajima2015information, nakajima2014exploiting}, have indicated that the interaction between the soft body and water is essential for computational tasks \cite{kagaya2022echo}. However, in our study, physical reservoir computing with soft tissue successfully performed computational tasks without an underwater environment. This could be attributed to the unique viscoelastic properties of soft tissues, which exhibit a power law dependence on time and frequency \cite{kobayashi2017simple, kobayashi2020non, bonfanti2020fractional}, leading to long term damping effects. 

These findings pave the way to successfully exploit the dynamics of human tissue for a wide range of computational tasks and applications. The miniaturization and flexibility of the sensors used in our method (for example, \cite{wang2022bioadhesive}), coupled with novel technology to implement linear readouts as devices, enables information processing using human tissue, eliminating the need for other computers  based on the Turing machine. Thus, this technology holds potential for a diverse range of applications such as human-machine interaction technology, medical devices, life support devices, and wearable devices.  Utilizing human tissue as a computational resource enables control and information processing to be implemented in targeted areas of the body, which could lead to the distribution of computational devices. It could also be applied to small-scale devices that perform information processing internally or in close proximity to the body, at the edge and locally. In addition, we expect our approach to inspire further research in novel concepts for computing across a wide range of fields. 

The framework presented in this study may also help shed light on the role of the human body and other living organisms. Physical reservoir computing can be regarded as an attempt to harness the inherent computational capability of living organisms. Exploring what kind of computational tasks can be performed using living bodies could provide a new perspective on understanding questions such as why living organisms possess their particular morphologies. There have been discussions concerning whether the body of a living organism could serve as a physical reservoir and function as a computational resource for perception and motor control of the organism \cite{nonaka2020locating}. The delay observed in the NARMA3 task, approximately 50 [ms], is comparable to the time required for reflection. Similarly, the delay of the NARMA10 system is approximately 150 [ms], which is comparable to the time taken for voluntary motor responses. These findings indicate that it is possible, from a timescale perspective, for soft tissue in living organisms to act as a physical reservoir and computational resource for perception and motor control. These insights highlight interesting research directions that warrant further exploration.

\begin{acknowledgments}
The authors would like to thank Y. Katagi for assistance with the experiments. This work was supported in part by a Grant-in-Aid for Scientific Research from the Ministry of Education, Culture, Sports, Science and Technology (MEXT) (19H02112 and 19K22878), Japan. 
\end{acknowledgments}

\appendix{
\section{\label{Preprocess} Data preprocessing}
Time series data of wrist joint angles were acquired at 1000 [Hz] using a goniometer. These data were resampled to 63 [Hz] to match the frame rate of an ultrasound image. A low-pass filter (Butterworth filter \cite{butterworth1930theory}) was applied to the data to remove noise. These data were used as the input joint angle vector \textit{$ \boldsymbol\theta= \left (\theta_{1}, \theta_{2}, \cdots   \right )$} where \textit{$\theta_{k}$} is the joint angle at k-th time data point. 
Ultrasound images of the wrist joint were acquired at 63[Hz] when the wrist was flexed and extended, and feature points in the ultrasound images were extracted by image processing at each time point. In determining the initial positions of the feature points, we first used Shi-Tomasi corner detection \cite{shi1994good} for the first frame of the ultrasound image. The coordinates of each feature point extracted by corner detection were tracked using optical flow \cite{lucas1981iterative} to obtain their x, y coordinate data. We used the coordinates of feature points that could be tracked to the end by optical flow \textit{$ \mathbf{\bar{s}} = \left (\bar{x}^{1},\bar{y}^{1}, \bar{x}^{2}, \bar{y}^{2}, \cdots   \right )^T $}, where (\textit{$\bar{x}^{i}$}, \textit{$\bar{y}^{i}$} ) is the coordinate of i-th feature point. Then, a standardization and low-pass filter (Butterworth filter \cite{butterworth1930theory}) was applied to the time series data of each coordinates \textit{$ \mathbf{\bar{s}}^i = (\bar{s}^{i}_1,\bar{s}^{i}_2, \cdots)$}. These data were used as the reservoir state vector \textit{$ \mathbf{s_k} = \left (s^{1}_k,s^{2}_k, s^{3}_k, s^{4}_k, \cdots   \right )^T$} at each time point k and the reservoir state matrix \textit{$\mathbf{S} (= \mathbf{s_{1}},\mathbf{s_{2}}, \cdots)$}. Note that the mean number of feature points used for the reservoir state vector was 995.
\section{\label{RidgeRegression} Ridge regression}
Ridge regression is a method used to calculate the weight \textit{$\mathbf{w}(=(w^{1},w^{2}, \cdots ))$} that minimizes the following loss function with the residual sum of squares term and the L2 regularization term. 
\begin{eqnarray}
L=\sum_{k}\left(o_k-\hat{o}_k\right)^2+\frac{1}{2} \lambda \sum_{i} (w^{i})^2
\label{eq:a1}
\end{eqnarray}
The weight \textit{$\mathbf{w}$}  that minimizes the loss function was determined analytically using the following equation.
\begin{eqnarray}
\mathbf{w}= \mathbf{o} \mathbf{S}^{\mathrm{T}} \left( \mathbf{S} \mathbf{S}^{\mathrm{T}}+\lambda \mathbf{I}\right)^{-1}
\label{eq:a2}
\end{eqnarray}
where \textit{$\mathbf{S}$} is the reservoir state matrix, which summarizes the reservoir state vectors described above \textit{$\mathbf{S} (=  \mathbf{s_{1}},\mathbf{s_{2}}, \cdots)$}, \textit{$\mathbf{w}$} is the weight of the corresponding linear readout, \textit{$\lambda $} is a hyperparameter and \textit{$\mathbf{I}$} is the identity matrix. The hyperparameter  \textit{$\lambda $} was determined by trial and error, and the same value (\textit{$\lambda=10$}) was used for all training. 
} 

\bibliographystyle{unsrt}
\bibliography{Manuscript} 

\end{document}